# Developing a Simple Hydrodynamic-based GIS-toolbox for Mapping the Suitable Zones for Aquatic Species Migratory

Mohamad Sarajzadeh[1*], Ehsan Tavakol bekhoda[2]

[1*] Department of Civil Engineering, Shahid Chamran University of Ahvaz, Ahvaz, Iran. Email: m-serajzadeh@stu.scu.ac.ir (*Corresponding author)

[2] Department of Civil Engineering, Islamic Azad University of Ahvaz, Ahvaz, Iran; Email: tavakol.ehsan@yahoo.com

**Abstract**

Several external sources may change the river's geomorphology, hydrodynamic characteristics, and water quality. Therefore, having a comprehensive assessment of the rivers' hydrodynamic would bring the preliminary information to reduce the human-made structures' short- and long-term effects on a different aspect of a region. An automated GIS toolbox is developed in the present study, which interpolates the hydrodynamic data from the HEC-RAS hydrodynamic model. A suitable fish habitat model is designed in the GIS toolbox, which uses the hydrodynamic interpolation result to assess the suitable habitat for aquatic species. In the study, different fish species' physical properties were considered to be used in conjunction with the hydrodynamic model to generate the habitat suitability raster datasets. This study provides the designers and scholars with a graphical interface of the GIS platform to model the potential migratory maps of the fish species under different hydrodynamic conditions.

**Introduction**

Recently, human-made construction activities have had different significant effects on the river's health. Construction activities such as bridges, dams, and culverts constructions are substantial activities that could change the water quality, leading to a mandatory migration for the aquatic species. In addition to the harmful effects of human-made construction activities on the water quality of the natural streams, in most cases, dams and culverts are blocking the migratory pass of the fishes swimming upstream of the rivers, especially in the spawning season (Katopodis 1992; Mahmoudian et al. 2019).

Fishways are the structures built on the sides of the dams or culverts to pass the fish upstream (FAO DVWK, 2002). As an effective solution to mitigate human-made structures on fish species, different fishways have been proposed, such as vertical slot fishway (VSF), Denil, pool and weir, etc. Therefore, the building of fishways is essential in hydraulic projects (Baharvand and Lashkar-Ara 2019; Song et al. 2019). The common goal of fishway structures in various studies is to adapt the conditions imposed on the ecosystem to create the optimal habitat condition compatible with the aquatic species (Baki et al. 2014; Rajaratnam et al. 1986, and 1992). Rajaratnam et al. (1986) presented extensive research on flow hydraulics in different configurations of VSF. Their research determined the formation of flow patterns and the conditions for creating flow patterns on average in seven different designs and four different geometric scales. Recently, Baharvand and Lashkar-Ara (2021) proposed a new meander fishway design (MMCF) used in steep natural streams (up to 20%) with a high rate of energy dissipation rate. Baki et al. (2020) investigated the hydraulic design aspects of rock-weir fishways with notch for habitat connectivity using numerical modeling. They identified two distinct flow regimes, weir and transitional. Many other studies have been done to assess the performance of the fishway structures under different hydraulic conditions (Baki et al. 2014; Rajaratnam et al. 1992).

In addition to the flow characteristics and the water quality, climate change and recent drought conditions could directly affect the water bodies (Hassanzadeh et al. 2020). Therefore, having an accurate prediction for the precipitation would bring valuable knowledge for water resources management to provide different approaches to mitigate the harmful environmental and ecological effects of drought and climate change. Some of the studies have considered the impact of industrialization and climate change in hydrology and water cycle, such as prediction of drought using learning-based approaches (Hassanzadeh et al. 2020), sediment yield and scouring estimation (Lashkarara et al. 2021), flood hazard systems (Puttinaovarat and Horkaew 2020).

A couple of methods are provided and developed by scientists and hydro-environmental agencies to assess the necessary equipment for measuring or predicting the hydrodynamic characteristics that affect aquatic life. Some studies have shown specific relationships between the aquatic species and their surrounding environment (Guisan and Zimmermann 2000; Joy and Death 2004). These kinds of species-environment models were used in different fields of water and ecology systems, including species conservation (Filipe et al. 2002; Wright et al. 1998), short and long-term flow regulation effects investigations (Joy and Death 2004; Marchant and Hehir 2002), and biological quality assessment (Brosse et al. 2001). Many studies have developed new models based on the GIS platform to study the species occurrence in conjunction with the multivariate statistical tools to connect different species and habitats. Besides, Artificial Neural Network was used with geospatial platforms to predict more accurate and reliable habitat models. It should be noted that the performance of learning-based in other water resources problems has been proved by different studies, such as assessment of hydraulic jump physical characteristics (Baharvand et al. 2020;

Roushangar et al. 2018), the bedload transport (Azamathulla et al. 2009), sediment scouring (Lashkar-Ara., 2021), investigation of flow characteristics in different spillways (Yildiz et al. 2020), etc.

Regardless of the soft computing techniques in water resources, hydrodynamic predictive models and GIS will be powerful tools to distinguish the sampled and unsampled sites (Peterson 2003). Nowadays, we face different hydro environmental problems resulting from industrialization and constructing the human-made structure on the river's basins. Therefore, having a general habitat model for modeling the safe zones for fish swimming paths would bring valuable information for having the best reaction to save the aquatic habitats. In this study, first, the GIS platform of the HEC-RAS hydrodynamic model will be discussed, and the required hydrodynamic result shapefiles (Velocity, Depth) will be exported. As the second phase of the study, python scripting will be used to develop an automated toolbox based on the predefined statistical libraries to generate the simulated river's low- and high-risk zones for migrating the aquatic species to the upstream of the case study. Developing an automated toolbox with the ability to read the exported hydrodynamic maps, interpolate and distinguish low and high-risk zones of a river based on the threshold of the aquatic species has not been studied yet. Besides, the model can assess the swimming capabilities of different fish species and generate each specific safe zones map. The result of the study will be produced as species-based classified shapefiles, which makes it easier to get the potential primary evaluation on a river basin for environmental assessments.

**Methodology and method**

*Ecological aspects*

Fish species can generate three different swimming speeds based on different situations as prolonged, sustained, and burst speed. Fishes have two different muscle types as red muscle and white muscle. The functionality of the mentioned muscles is wholly separated (Behlk 1991). Red muscles produce the appropriate energy for long-term swimming using prolonged or sustained swimming speed (Katopodis and Gervais 2012). White forces can produce a high energy rate for fish swimming, although the generated energy can be used for short-term swimming and very high speed (burst speed). Greek-Walker (1975) proved that the white muscle produced energy is the red muscles produced about four times the energy for cold-water fish. Therefore, the total energy which was produced by the fishes' muscles should be considered in terms of the biological power to overcome the accessible hydrodynamic power in the pools and outgoing jet at the slots.

*Biologic power*

Furnis et al. (2006) introduced that the flow velocity and depth of the water are the main variables affecting the ability of the fishes to overcome the current to migrate towards the upstream of the structure. Equation (1) and (2) shows the hydraulic and biological power of fishes, respectively; The superiority of the biological power released from fish species to the hydraulic power of the structure pools is the primary indicator of the fish's ability to migrate upstream of the structure. The velocity of water plays the most critical role (Behlk 1991).

$$PW = \gamma V S_0 \qquad (1)$$

$$PW_f = P_f V_{bf} \qquad (2)$$

where *V* is the mean flow velocity, *P.W.* is hydraulic power in each pool, *PW$_f$* is the biological power of the fish using burst speed, and *V$_{bf}$* is the velocity of the fishes' body generated by the white muscles.

*Fish species of the study*

Oncorhynchus mykiss is one of the essential fishes found in cold-water tributaries in North America and Asia. Akanyeti and Liao (2013) studied the swimming speed of the rainbow trout. Their result was another proof to other research done to identify the influential variables of fish swimming. Fish swimming ability is directly connected to the length of the fish and time to exhaustion. Table (1) shows a list of essential fishes' swimming speeds. The velocity of each fish species could be calculated using equation (3).

$$V_f = aL^b t^c \tag{3}$$

where; $V_f$ is the fish swimming speed, L is the fish body length, t is the time to exhaustion. In addition, a, b, and c are regression constants based on fish species. Figure (1) shows different body lengths of the fish, which should be used in appropriate conditions based on the information considered in Table (1) for swimming speed calculations.

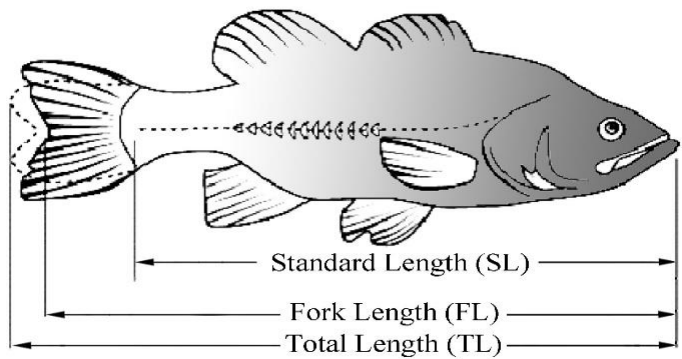

**Figure 1.** Different lengths of the fishes' body

Standard length is rarely used for swimming speed calculation. Table (1) shows that fork and total sizes are commonly used length parameters for various fish species. For Oncorhynchus mykiss fish (Rainbow trout), the total length should be taken in the fish speed equation (7), and constant regression coefficients are 12.81, 1.07, and 0.4, respectively.

**Table 1.** Swimming capabilities for some populated fish species (Furnis et al. 2006)

| Fish species | Length type | Burst speed coefficient | | | Prolonged speed coefficient | | | Body length | | Time to exhaustion (s) | |
|---|---|---|---|---|---|---|---|---|---|---|---|
| | | a | b | c | a | b | c | Min | Max | Burst | Prolong |
| Archetic char | TL | 4.30 | 0.00 | 0.49 | 2.69 | 0.606 | 0.080 | 70 | 420 | 18 | 1800 |
| Arctic Grayling | FL | - | - | - | 1.670 | 0.193 | 0.100 | 60 | 400 | - | 600 |
| Atlantic Salmon | TL | 11.34 | 0.88 | 0.50 | 0.173 | 0.680 | 0.500 | 52 | 500 | 10 | 1800 |
| Broad Whitefish | FL | - | - | - | 1.460 | 0.450 | 0.100 | 50 | 400 | - | 600 |
| Brook Trout | TL | - | - | - | 1.990 | 0.430 | 0.100 | 40 | 270 | - | 4500 |
| Burbot | FL | - | - | - | 2.230 | 0.070 | 0.26 | 100 | 700 | - | 600 |
| Brown Trout | TL | 8.74 | 0.68 | 0.50 | - | - | - | 128 | 373 | 2 | - |
| Coho | TL | 13.30 | 0.52 | 0.65 | - | - | - | 256 | 610 | 10 | - |
| Dace | TL | 12.37 | 0.65 | 0.5 | - | - | - | 30 | 250 | 20 | - |
| Flathead Chub | FL | - | - | - | 2.660 | 0.670 | 0.100 | 150 | 350 | - | 600 |
| Goldfish | TL | 5.37 | 0.66 | 0.22 | - | - | - | 67 | 213 | 20 | - |
| Humpback Whitefish | FL | - | - | - | 1.730 | 0.350 | 0.100 | 60 | 600 | - | 600 |
| Inconnu | FL | - | - | - | 1.290 | 0.175 | 0.100 | 70 | 800 | - | 600 |
| Longnose Sucker | FL | - | - | - | 2.390 | 0.529 | 0.100 | 30 | 700 | - | 600 |
| Northern Pike | FL | - | - | - | 1.170 | 0.550 | 0.100 | 100 | 800 | - | 600 |
| **Rainbow trout** | **TL** | **12.81** | **1.07** | **0.48** | **-** | **-** | **-** | **103** | **813** | **10** | **-** |
| Sea Lamprey | TL | - | - | - | 2.570 | 0.36 | 0.26 | 145 | 508 | - | 1635 |
| Sockeye | TL | - | - | - | 5.470 | 0.890 | 0.070 | 126 | 611 | - | 1800 |
| Steelhead | TL | 12.81 | 1.07 | 0.48 | - | - | - | 103 | 813 | 10 | - |
| Walleye | FL | - | - | - | 2.600 | 0.510 | 0.100 | 70 | 400 | - | 600 |
| White Sucker | FL | - | - | - | 2.480 | 0.552 | 0.100 | 10 | 400 | - | 600 |

*Case study and data used*

The case study was chosen as a section of the Bald Eagle Creek downstream of the Foster Joseph Sayers Reservoir, Pennsylvania, USA. Foster Joseph Sayers Reservoir is a 1,700-acre lake in the central part of Pennsylvania. Figure (2) shows the location of the case study. As illustrated in Figure (2), the study area lies between latitude 77º36'30" W to 77º36'30" W and longitude -41º03'00" N to -41º03'30" N. According to the National Hydrography Dataset high-resolution flowline data of USGS, the average daily mean flow discharge on August 2020 was reported as 20 m$^3$/s (at the initial condition section of the 2D model). Due to a lack of bathymetric data for this study, the existing DEM was picked as the bathymetry corrected file.

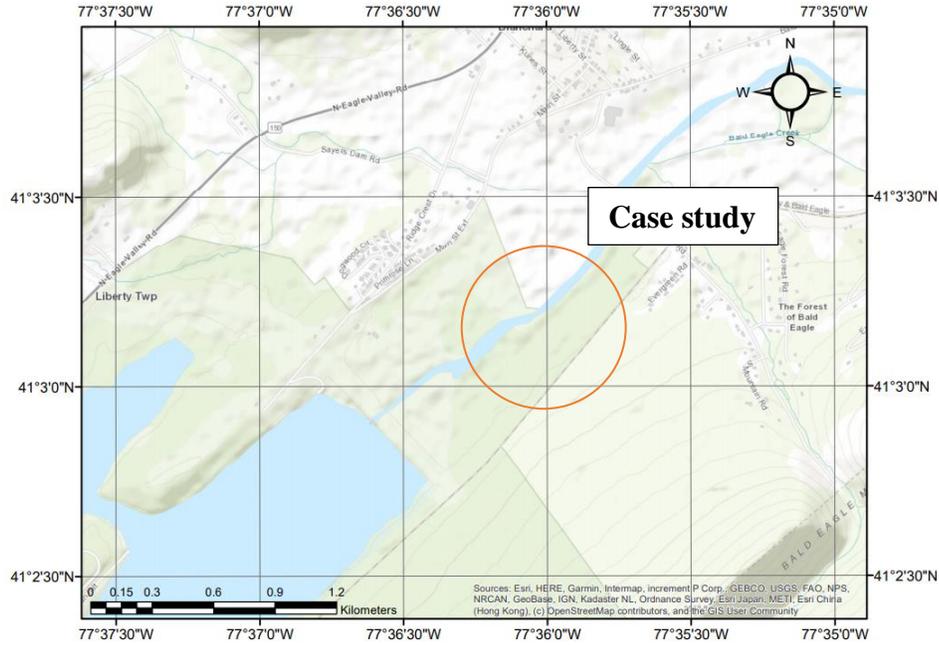

**Figure 2.** Location of the case study at Bald Eagle Creek downstream of the Foster Joseph Sayers Reservoir, Pennsylvania, USA

*Data used and hydrodynamic model preparation*

Measuring water in open channels is an essential component of water management and water resources problems (Samani et al. 2021). Flow discharge has a significant role in fish and aquatic studies (Rajaratnam et al. 1986). Therefore, accurate hydrodynamic modeling is required to numerically model the current condition based on the geomorphology of the river. There are a couple of numerical models in one and two-dimensional systems to compute the flow characteristics. Most well-known models are working based on the St. Venant equations such as HEC-RAS, MIKE 11, etc. The HEC-RAS is a widely used hydrodynamic model which the U.S. Army Corps of Engineers develop. The first version of the model was publicly released in 1995 in one-dimensional space. In addition, HEC-RAS was used by different researchers and agencies for flood inundation modeling and dam-break modeling (Downs et al., 2009). In this study, HEC-RAS 2D was used as the primary hydrodynamic model. However, any hydrodynamic model that can generate the raster dataset can be used to input the developed habitat model in this study. Digital Elevation Model (DEM) with 10 m resolution was used to model the channel's geomorphology of the channel. The streamflow and gage height datasets were exported from the USGS gauge station at the site location. The architecture of the developed fish habitat suitability model is indicated in Figure (3).

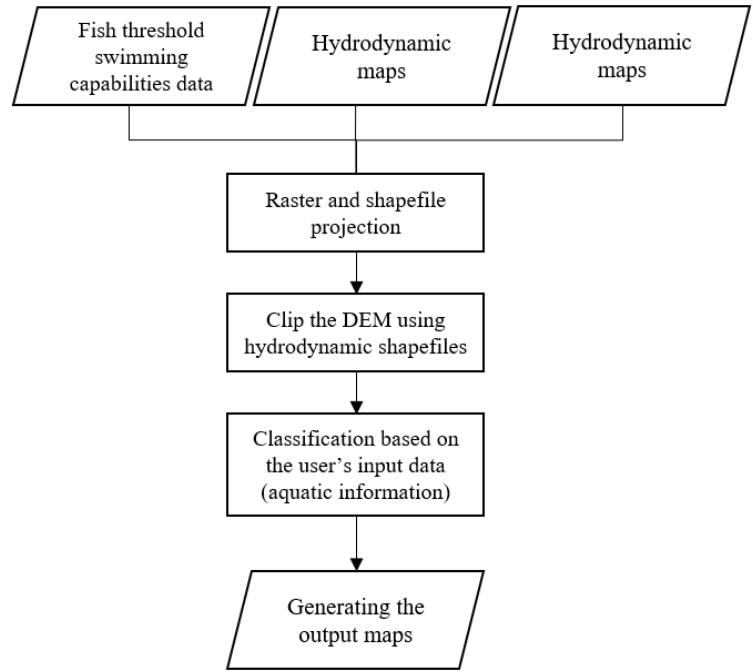

**Figure 3**. The architecture of the GIS Toolbox developed for mapping the safe zones

A sensitivity analysis test was applied to the mesh size of the two-dimensional mesh domain of the HEC-RAS 2D model to find the optimal value of the mesh grid sizes. Figure (4) shows the total number of mesh cells is 1133, and the manning number used as 0.06 uniformly for the entire over banks and main channel.

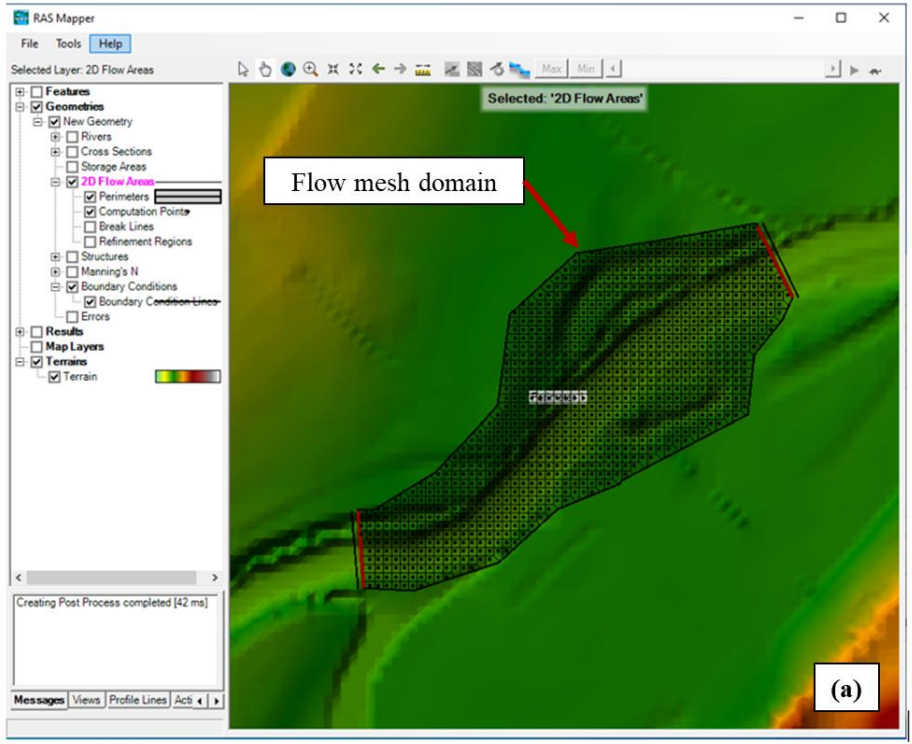

**Figure 4.** Mesh domain and manning number of the case study

Depth-averaged velocity and flow depth raster datasets were exported as the input variables to the developed fish habitat suitability model. The spatial analysis and statistics were used to generate the habitat model discussed in the next section.

*Spatial Analysis Methods*

Interpolation, Extraction, and Reclassification are the main spatial analysis methods that make the basis of this research. These methods will be used in conjunction with the other GIS toolboxes using the Arcpy package inside the python. IDW and Spline interpolation is added to the model as two interpolating options to the clients. In addition, the user will be able to change the reclassification criteria based on the project inside the toolbox. Figure (5) shows the main window of the Graphical User Interface of the created GIS toolbox to predict the suitable zones for different fish species in natural streams.

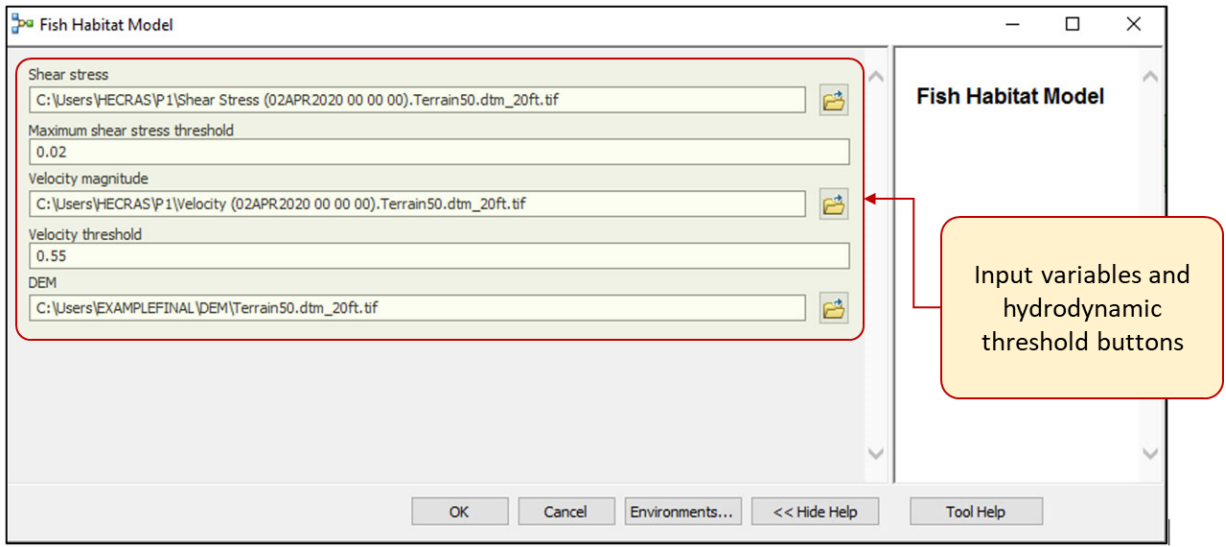

**Figure 5.** The main menu of the GIS toolbox of fish habitat suitability modeling

Based on the goal of this project, which is developing an automated toolbox for mapping safe migratory zones for fish species as a part of ArcMap Command toolboxes, two different ways could be used for creating the toolbox structure. Because of the requirements of this project, the base structure of the model was built based on the python code. Figures (6) and (7) show the preprocess and main process code of the developed model in python, respectively.

```python
# Import arcpy module
import arcpy

# Check out any necessary licenses
arcpy.CheckOutExtension("spatial")

# Script arguments
Shear_stress = arcpy.GetParameterAsText(0)
if Shear_stress == '#' or not Shear_stress:
    Shear_stress = "C:\\Users\\HECRAS\\P1\\Shear Stress (02APR2020 00 00 00).Terrain50.dtm_20ft.tif" #

Maximum_shear_stress_threshold = arcpy.GetParameterAsText(1)
if Maximum_shear_stress_threshold == '#' or not Maximum_shear_stress_threshold:
    Maximum_shear_stress_threshold = "0.02" # provide a default value if unspecified

Velocity_magnitude = arcpy.GetParameterAsText(2)
if Velocity_magnitude == '#' or not Velocity_magnitude:
    Velocity_magnitude = "C:\\Users\\HECRAS\\P1\\Velocity (02APR2020 00 00 00).Terrain50.dtm_20ft.tif"

Velocity_threshold = arcpy.GetParameterAsText(3)
if Velocity_threshold == '#' or not Velocity_threshold:
    Velocity_threshold = "0.65" # provide a default value if unspecified

DEM = arcpy.GetParameterAsText(4)
if DEM == '#' or not DEM:
    DEM = "C:\\Users\\DEM\\Terrain50.dtm_20ft.tif" # provide a default value if unspecified
```

**Figure 6.** The initial section of the habitat model python reference code (input necessary variables into the model)

According to figure (6), the model first reads the input data from the hydrodynamic model. The maximum shear stress and velocity threshold will be introduced to the model as the main hydrodynamic parameters affecting the aquatic habitat inside the case study. The model will process the suitability index of the stream based on the ecological aspects considered based on the local fish species.

Figure (7) illustrates the program reference code used to generate the spatially distributed effective data. Based on Figure (7), the developed fish habitat model will assess the hydrodynamic condition. After clipping the safe zones for any specific fish species inside the study area, reclassification and extraction commands were used as the following tools for creating the shapefiles addressing the safe zones according to each fish species available in each case study.

```
# Local variables:
Shear_calculated = Shear_stress
Reclassed_shear = Shear_calculated
Shear_masked = Reclassed_shear
Depth_masked = Shear_masked
Property = Depth_masked
Velocity_SHP__3_ = Property
Velocity_SHP__6_ = Velocity_SHP__3_
Velocity_SHP__5_ = Velocity_SHP__6_
Velocity_Calculated = Velocity_magnitude
Reclassed_velocity = Velocity_Calculated
Dem_masked = Reclassed_velocity
Velocity_SHP = Reclassed_velocity
Velocity_SHP__2_ = Velocity_SHP
Depth = "C:\\Users\\HECRAS\\P1\\Depth (02APR2020 00 00 00).Terrain50.dtm_20ft.tif"

# Process: Raster Calculator (2)
arcpy.gp.RasterCalculator_sa("\"%Velocity magnitude%\" <= float(%Velocity threshold%)", Velocity_Calculated)

# Process: Reclassify
arcpy.gp.Reclassify_sa(Velocity_Calculated, "Value", "1 1", Reclassed_velocity, "NODATA")

# Process: Extract by Mask
arcpy.gp.ExtractByMask_sa(DEM, Reclassed_velocity, Dem_masked)
# Process: Raster to Polygon
arcpy.RasterToPolygon_conversion(Reclassed_velocity, Velocity_SHP, "SIMPLIFY", "VALUE")
# Process: Add Field
arcpy.AddField_management(Velocity_SHP, "MeanDepth", "FLOAT", "", "", "", "", "NULLABLE", "NON_REQUIRED", "")
# Process: Raster Calculator
arcpy.gp.RasterCalculator_sa("\"%Shear stress%\" <= float(%Maximum shear stress threshold%)", Shear_calculated)
# Process: Reclassify (2)
arcpy.gp.Reclassify_sa(Shear_calculated, "Value", "0 1 1", Reclassed_shear, "NODATA")
# Process: Extract by Mask (2)
arcpy.gp.ExtractByMask_sa(Reclassed_shear, Reclassed_velocity, Shear_masked)
# Process: Extract by Mask (3)
arcpy.gp.ExtractByMask_sa(Depth, Shear_masked, Depth_masked)
# Process: Get Raster Properties
arcpy.GetRasterProperties_management(Depth_masked, "MEAN", "")
# Process: Calculate Field
arcpy.CalculateField_management(Velocity_SHP__2_, "MeanDepth", Property, "VB", "")
# Process: Add Field (2)
arcpy.AddField_management(Velocity_SHP__3_, "Volume", "FLOAT", "", "", "", "", "NULLABLE", "NON_REQUIRED", "")
# Process: Calculate Field (2)
arcpy.CalculateField_management(Velocity_SHP__6_, "Volume", "[MeanDepth] * [Shape_Area]", "VB", "")
```

(Import necessary hydrodynamic and ecological variables | Spatial calculation process of the developed fish habitat suitability model)

**Figure 7.** The main process section of the habitat model python reference code

The potential volume of the safe zones will be calculated using the average water flow depth in the safe zone areas. The result of the habitat model considers the total volume of water where the fish species could have in their path to the upstream using the prolonged speed. Also, due to the burst speed thresholds, the possibility of dangerous areas could be studied as well. The result of the model for three different fish species will be discussed in the following section.

**Result and discussion**

Three different fish species with three different velocity thresholds have been used to model the safe zones for Bald Eagle Creek. Table (2) shows the assigned shear stress and velocity thresholds to the model. The maximum capacity for shear stress remains constant due to a lower effect on the migratory fish process. Also, the velocity threshold was changed as it is the most important parameter affecting the fish swimming behavior.

**Table 2.** The assumed fish velocity and shear stress threshold

| Fish species | Velocity (m/s) | Shear stress (N/m$^2$) |
|---|---|---|
| Fish no. 1 | 0.65 | |
| Fish no. 2 | 0.45 | 0.02 |
| Fish no. 3 | 0.35 | |

Figure (8) shows the habitat suitability map for the fish category 1. Because of the high-velocity threshold comparing to both other fish species (groups 2 and 3) is expected to have a larger safe zone area. As can be seen, the total safe area for fish groups 2 and 3 resulted in a smaller area.

Besides the potential migratory maps for fishes, the attribute tables of the habitat model will provide a significant result of the area, including the possible safe volume of flow (based on the mean depth of the water along the basin) and the average streamwise length of each safe zones.

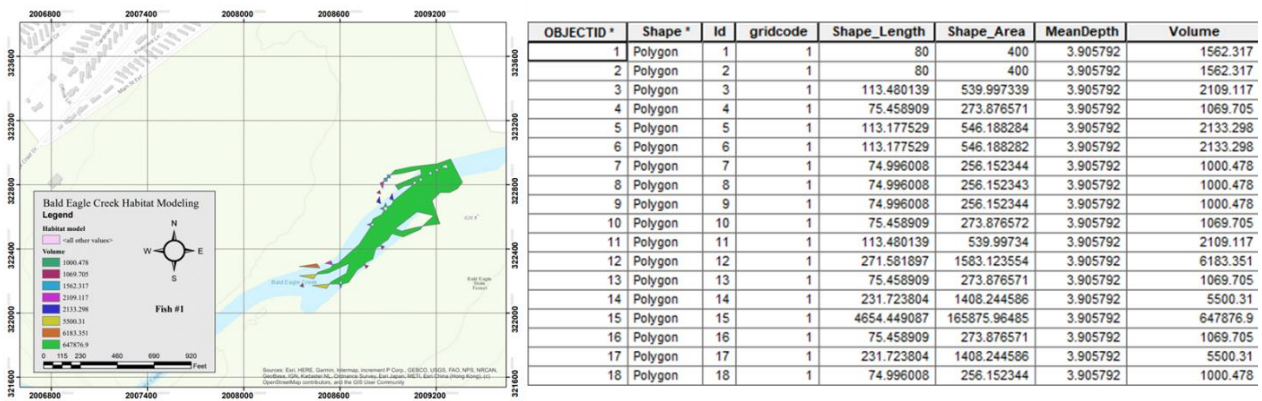

**Figure 8.** Potential migratory safe zones for fish species group #1

Figures (9) and (10) show the result of a habitat model for fish species groups 2 and 3 for the discharge of 20 m$^3$/s.

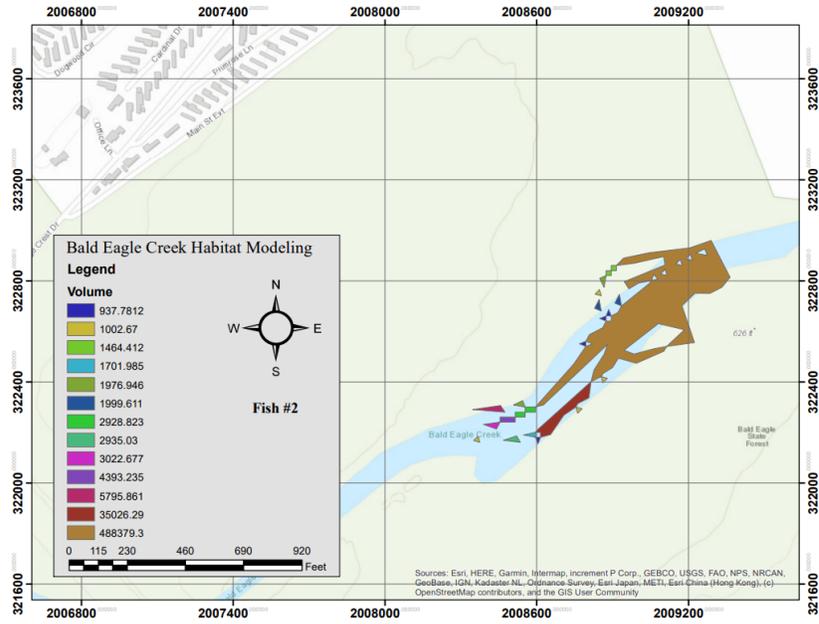

**Figure 9.** Potential migratory safe zones for fish species group #2

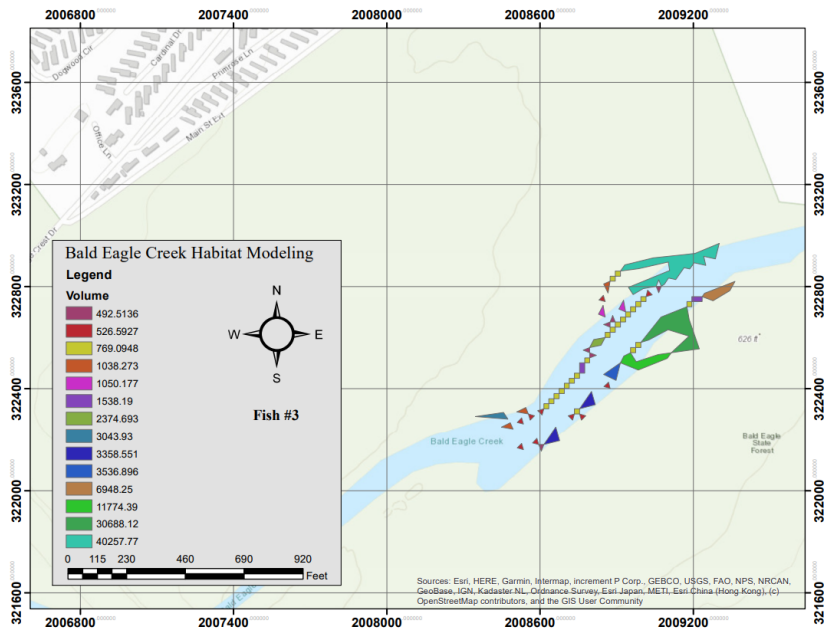

**Figure 10.** Potential migratory safe zones for fish species group #3

**Conclusion**

In this study, the relationship between the fish species swimming capabilities and hydrodynamic model results was used to find suitable and safe zones for fish swimming upstream without any potential prevention due to the hydrodynamic result. In this study, the full structure was done by python 2.7, and the "arcpy" package has been used widely to produce the appropriate results' shapefiles. Digital Elevation Model (DEM), Shear stress raster map, Velocity raster map are assigned as the main hydrodynamic parameters to the model. The threshold of the shear stress and velocity of the fish species should be introduced to the model as the required datasets.

In the present study, the designed fish habitat model will generate the geospatial maps of the safe zones for fish migratory to upstream. In this report, the result of the Bald Eagle Creek hydrodynamic model was used for three different fish swimming capacities. The model gives the safe area and the potential total safe volume, which could be used to investigate the number of fish swimming upstream. Having an estimation of a possible number of migratory fishes could be used to mitigate the limitations in their path and provide some management plan to increase the safe zones for fishes in exhaustion time.